\documentclass[aps, prd, twocolumn, superscriptaddress, nofootinbib]{revtex4}
\usepackage{graphicx}% Include figure files
\usepackage{dcolumn}% Align table columns on decimal point
\usepackage{amssymb}% outlined letters
\usepackage{amsmath,amssymb,amsfonts}
\usepackage{amsmath}
\usepackage{bbm}
\usepackage{hyperref}

\begin{document}

%Macros
\newcommand{\Eq}[1]{\mbox{Eq. (\ref{eqn:#1})}}
\newcommand{\Fig}[1]{\mbox{Fig. \ref{fig:#1}}}
\newcommand{\Sec}[1]{\mbox{Sec. \ref{sec:#1}}}

\newcommand{\PHI}{\phi}
\newcommand{\PhiN}{\Phi^{\mathrm{N}}}
\newcommand{\vect}[1]{\mathbf{#1}}
\newcommand{\Del}{\nabla}
\newcommand{\unit}[1]{\;\mathrm{#1}}
\newcommand{\x}{\vect{x}}
\newcommand{\ScS}{\scriptstyle}
\newcommand{\ScScS}{\scriptscriptstyle}
\newcommand{\xplus}[1]{\vect{x}\!\ScScS{+}\!\ScS\vect{#1}}
\newcommand{\xminus}[1]{\vect{x}\!\ScScS{-}\!\ScS\vect{#1}}
\newcommand{\diff}{\mathrm{d}}

\newcommand{\be}{\begin{equation}}
\newcommand{\ee}{\end{equation}}
\newcommand{\bea}{\begin{eqnarray}}
\newcommand{\eea}{\end{eqnarray}}
\newcommand{\vu}{{\mathbf u}}
\newcommand{\ve}{{\mathbf e}}

%=====================================================================
%=====================================================================
%=====================================================================

\title{Planck-scale phenomenology with anti-de Sitter momentum space}

\newcommand{\addressImperial}{Theoretical Physics, Blackett Laboratory, Imperial College, London, SW7 2BZ, United Kingdom}
\newcommand{\addressRoma}{Dipartimento di Fisica, Universit\`a “La Sapienza”
and Sez. Roma1 INFN, P.le A. Moro 2, 00185 Roma, Italia}

\author{Michele Arzano}
\affiliation{\addressRoma}
\author{Giulia Gubitosi}
\affiliation{\addressImperial}
%\affiliation{\addressRoma}
\author{Jo\~{a}o Magueijo}
%\email{magueijo@ic.ac.uk}
\affiliation{\addressImperial}
%\affiliation{\addressRoma}
\author{Giovanni Amelino-Camelia}
\affiliation{\addressRoma}

\date{\today}

\begin{abstract}
We investigate the anti-de Sitter (AdS) counterpart to the well studied de Sitter (dS) model for energy-momentum space, viz ``$\kappa$-momentum space'' space (with a structure based on the properties of the $\kappa$-Poincar\'e Hopf algebra). On the basis of previous preliminary results one might expect the two models to be ``dual'': dS exhibiting an invariant maximal spatial momentum but unbounded energy, AdS a maximal energy but unbounded momentum. If that were the case AdS momentum space could be used to implement a principle of maximal Planck-scale energy, just as several studies use dS momentum space to postulate of maximal Planck-scale spatial momentum. However several unexpected features are uncovered in this paper, which limit the scope of the expected duality, and interestingly they take different forms in different coordinatizations of AdS momentum space.  ``Cosmological'' AdS coordinates mimic the dS construction used for $\kappa$-momentum space, and produce a Carrol limit in the ultraviolet. However, unlike the $\kappa$-momentum space, the boundary of the covered patch breaks Lorentz invariance, thereby introducing a preferred frame. In ``horospherical'' coordinates we achieve full consistency with frame independence as far as boost transformations are concerned, but find that rotational symmetry is broken, leading to an anisotropic model for the speed of light.  Finally, in ``static'' coordinates we find a way of deforming relativistic transformations that successfully enforces frame invariance and isotropy, and produces a Carrol limit in the ultraviolet. However, the phenomenological implications appear to be too weak for any realistic chance of detection. Our results are also relevant for a long-standing debate on whether or not coordinate redefinitions in momentum space lead to physically equivalent theories: our three proposals are evidently physically inequivalent, leading to alternative models of Planck-scale effects. As a corollary we study the UV running of the Hausdorff dimension of momentum space in the first and third model, obtaining different results.
\end{abstract}


\maketitle

%=====================================================================
%=====================================================================
%=====================================================================

\section{Introduction}
In recent years findings in several areas of quantum-gravity research (see, {\it e.g.}, Refs.~\cite{AmelinoCamelia:2008qg,Mattingly:2005re} and references therein)  have motivated  the investigation of Planck-scale modified dispersion relations (MDRs), and this has attracted interest in MDRs as a possible avenue for Planck-scale phenomenology associated with astrophysical and cosmological observations \cite{AmelinoCamelia:1997gz, Jacob:2006gn, Alexander:2001dr, Alexander:2001ck, Gubitosi:2009eu}. It has become clear that some of the key predictions arising from MDRs depend crucially on whether the relevant framework {\it breaks} or merely {\it deforms} relativistic symmetries. A preferred-frame scenario is inevitable if the transformation laws between inertial observers remain the standard special-relativistic ones, since they only leave invariant the usual Einsteinian dispersion relation
$E^2 -p^2 =m^2$.  However, it is possible to introduce a deformation of relativistic symmetries preserving the equivalence of reference frames and leaving the MDR observer-independent \cite{AmelinoCamelia:2005ne, Magueijo:2001cr, KowalskiGlikman:2002jr}. Notable examples of such ``DSR" (doubly-special, or deformed-special relativity) scenarios include theories based on a maximally-symmetric curved momentum space. This has been investigated  in great detail if momentum space has de Sitter geometry. Here we seek to investigate momentum space with anti-de Sitter geometry, a possibility which has so far received very little attention in the literature (see, however, \cite{AmelinoCamelia:2012rz,Arzano:2014ppa}). In doing so we will uncover several significant differences between dS and AdS models of momentum space.

If DSR-relativistic scenarios arise from maximally-symmetric momentum space it is easy to see how one can achieve compatibility between some MDRs and the laws of transformation between inertial observers.  One usually introduces ordinary special relativity by taking as the starting point the isometries of Minkowski spacetime, but one could equally well start from the isometries of Minkowski momentum space. In either case one can derive the transformation laws of momenta and spacetime coordinates by consistency \cite{principle}.  Since the isometries of de Sitter (or anti-de Sitter) space can be seen as a deformation of the isometries of Minkowski space, any set of transformation laws derived from the isometries of de Sitter (or anti de Sitter) momentum space  is as ``relativistic'' as special relativity (i.e. it abides by the principle of the relativity of inertial frames). However, such a construction entails a deformations of the transformation laws between inertial observers, and these will leave invariant a modified (deformed) dispersion relation.

Constructions based on de Sitter momentum space have been extensively studied in the literature, with many authors registering the expectation that the counterpart AdS model would have properties easily obtainable from those of dS momentum space. Several arguments suggest that the two models should be ``dual'', with dS exhibiting an invariant maximal spatial momentum but unbounded energy, and AdS a maximal energy but unbounded spatial momentum. However, as we will show in this paper, many crucial novelties arise in AdS curved momentum space
that are not captured by this expected duality. Whereas previous arguments focused exclusively on local properties of the two momentum spaces, one of the key ingredients of our analysis is the realization of the fact that different coordinates cover different patches of the manifold, and that this leads to different physical statements on what is the free theory. A number of options appear, mimicking---or not---constructions previously considered for dS. We will find that in all of them AdS momentum space is {\it qualitatively} very different from dS, the main point made in this paper.

An important reference for us is the so-called ``$\kappa$-momentum space", a coordinatization of a certain patch of de Sitter momentum space which has been found to have remarkably good relativistic properties, and can be inspired by the formal structure of the $\kappa$-Poincar\'e Hopf algebra \cite{Lukierski:1992dt, Lukierski:1993wxa, Lukierski:1991pn}. As we observed in Ref.\cite{dsrrsd}, $\kappa$-momentum space can be viewed as the momentum space equivalent of the ``cosmological'' representation of dS spacetime. After reviewing this construction of $\kappa$-momentum space (Section~\ref{dSbackground}) in Section~\ref{cosmo} we find the corresponding construction for AdS.  In  such ``cosmological'' coordinates a simple representation for the Casimir invariant, momentum space metric and integration measure is found. However, unlike with dS, the boundary of the covered patch breaks Lorentz invariance. If analyzed only at the level of infinitesimal transformations the model is DSR-relativistic, but a breakdown  of Lorentz invariance is noticed when considering finite Lorentz transformations. We know of no previous examples in the literature of such subtle breakdown of relativistic symmetries (see, however \cite{Magueijo:2002am}), and we speculate that such a possibility could play an important role in the phenomenology of departures from ordinary special relativity, a case somewhere in between the one of full breakdown of relativistic symmetries (breakdown appreciable already for infinitesimal transformations) and the DSR-relativistic case (fully relativistic picture).

Another possible approach mimicking the $\kappa$-momentum space consists of using  ``horospherical'' coordinates, which cover a patch of AdS. We do this in Section~\ref{horo}, only to encounter a similar problem to that found for cosmological coordinates, but this time regarding the rotations. Similarly to what happens for boosts in ``cosmological" coordinates, the boundary of the patch covered by horospherical coordinates breaks invariance under rotations, and so the theory is anisotropic. The ensuing formalism is somewhat awkward, and the expression for the Casimir is far more complex. However, we argue that this could be a good model for encoding anisotropic MDRs and speed of light. We should however bear in mind some potential pathologies: the model does not allow one spatial momentum to take arbitrary negative values if we want to preserve invariance under finite boosts.

In view of the symmetry breaking properties of these two models, in Section~\ref{static} we investigate an alternative construction which does not purport to mimic the $\kappa$-momentum space.  We propose a set of coordinates analogous to ``static'' coordinates in the spacetime picture. They cover the whole of AdS and do not break Lorentz invariance in any way. They lead to simple expressions for the metric, Casimir and integration measure. As with the first model, we find a Carroll limit in the UV, i.e.: the speed of light goes to zero in the UV.

As an application,  in Section~\ref{run} we briefly investigate the issue of running of dimensionality for the first and third model (the matter is far less obvious for the second model, due to its anisotropy). We do this by choosing linearizing coordinates and evaluating the {\it measure of integration on momentum space} (a procedure described in~\cite{measure,dsrrsd}, known to match the spectral dimension in the UV limit in all cases studied so far).  We find that the two models exhibit running to different dimensions, a particularly transparent indication
of the fact that they are physically distinct models, though both based on AdS momentum space.

Given that static coordinates have not been considered for dS, for completeness in Section~\ref{desitter} we present them. We find that they break Lorentz invariance in a fashion similar to that found for AdS in cosmological coordinates.  We also examine running of dimensionality in the corresponding model, finding a very suggestive result.  In a concluding Section we collect the main results of this paper and discuss their implications.

\section{de Sitter momentum space}\label{dSbackground}
As mentioned in the Introduction the action of relativistic symmetries on momenta can be {\it deformed} if one considers a maximally symmetric curved momentum space. A widely studied example of deformed Poincar\'e symmetries reflecting such non-trivial geometry of momentum space is the so-called $\kappa$-Poincar\'e algebra \cite{Lukierski:1992dt, Lukierski:1993wxa, Lukierski:1991pn}. Indeed, as first shown in \cite{KowalskiGlikman:2004tz}, in a $\kappa$-deformed framework momenta can be seen as coordinates on a portion of de Sitter momentum space defined as a four-dimensional hyper-surface:
\begin{equation}\label{4}
  -P_0^2 + P_1^2 + P_2^2 + P_3^2 + P_4^2 =\frac{1}{\ell^2}\, .
\end{equation}
embedded in five dimensional Minkowski space, with line element:
\be
ds^{2}=-dP_{0}^{2}+d P_{1}^{2}+d P_{2}^{2}+d P_{3}^{2}+d P_{4}^{2}\, ,
\ee
selected by the inequality
\begin{equation}
P_0-P_4>0\,,
\end{equation}
where the ``cosmological constant" is the inverse of the parameter which governs the deformation of the algebraic structures in $\kappa$-Poincar\'e, $\kappa=1/\ell$. The natural parameterization of this submanifold, inherited by the bi-crossproduct basis of the $\kappa$-Poincar\'e algebra \cite{MajidRuegg}, is given by  {\it bi-crossproduct coordinates}, which correspond in position space to the ``cosmological'' or ``flat slicing" rendition of de Sitter space. They  are related to the embedding coordinates via:
\begin{eqnarray}\label{bicrossp}
 {P_0}(E, \vec{p}) &=&  \frac{\sinh
(\ell {E})}{\ell} + \frac{\ell p^2}{2}\,
e^{  \ell E}, \nonumber\\
 P_i(E, \vec{p}) &=&  - p_i \, e^{\ell E}, \nonumber\\
 {P_4}(E, \vec{p}) &=&  -\frac{\cosh
(\ell E)}{\ell} + \frac{\ell p^2}{2}\, e^{\ell E},
\end{eqnarray}
where $p\equiv|\vec{p}|$.
With these coordinates the line element takes the familiar ``cosmological'' de Sitter metric form:
\be
ds^2
=- dE^2+e^{2\ell E}\sum^{3}_{j=1}dp_j^2
\ee
from which it is easy to infer the invariant integration measure in momentum space:
\be\label{measurep}
d\mu(E,\vec p)=e^{3 \ell E} p^{2} dE dp\; .
\ee

\begin{figure}
\includegraphics[scale=0.5]{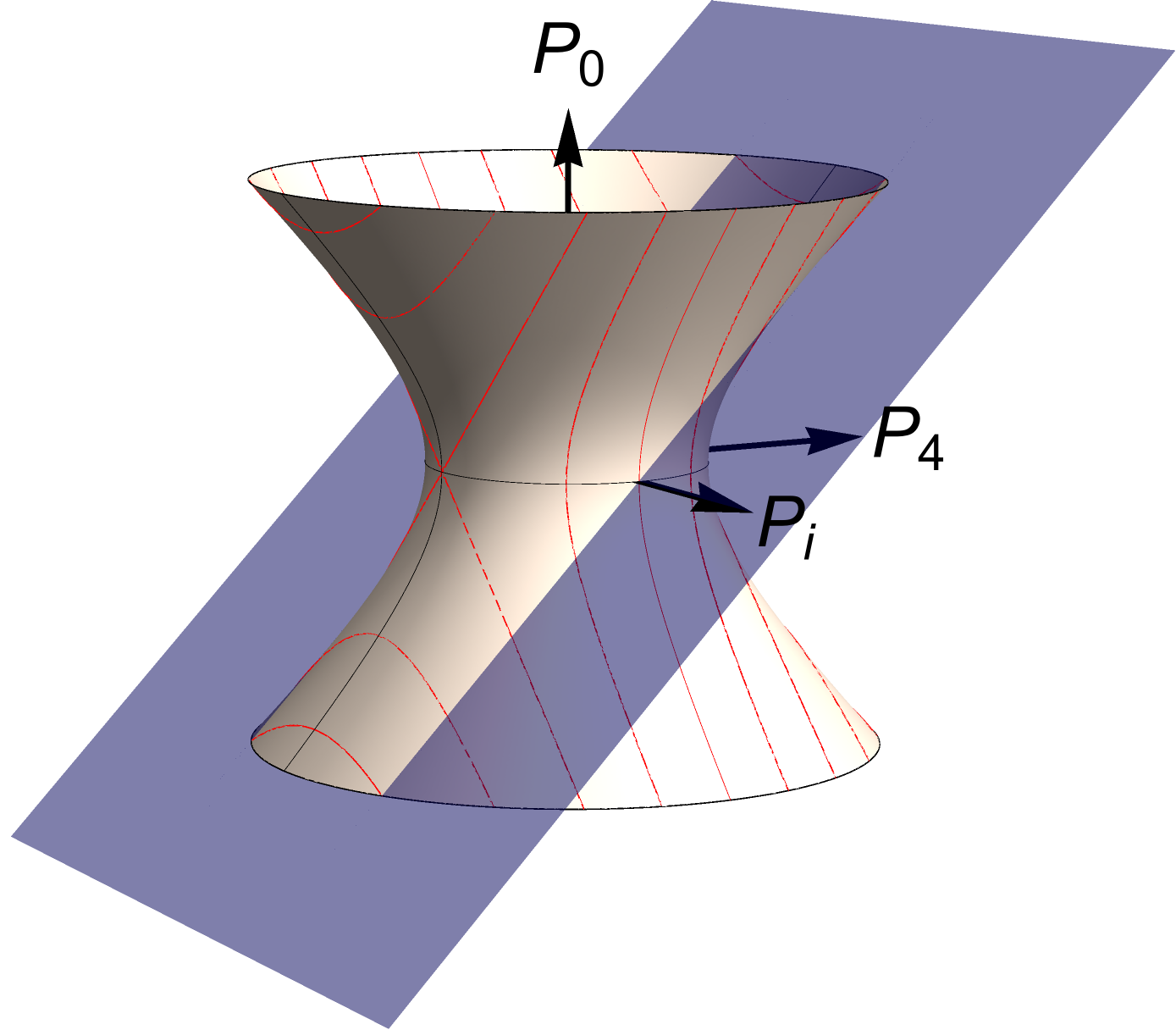}
\caption{The portion of (2-dimensional) de Sitter momentum space known as the ``2D $\kappa$-mometum space".
The blue plane is defined by the condition $P_0-P_4=0$, so  the $\kappa$-mometum space
is on the upper-left side of the plane. The red lines represent the mass-shells, defined by the condition $P_4=const$. In order to have the mass-shells completely within the allowed region one has to further
restrict to $P_0 > 0$ and $P_4< 0$.}
\label{fig:dSbicross}
\end{figure}

The deformed mass-shell is given by the intersection of a plane $P_4=const.$  with the momentum manifold:
\be\label{mass-shell}
  -P_0^2 + \vec{P}^2=\frac{1}{\ell^2}-P_4^2=m^2\, .
\ee
In Figure \ref{fig:dSbicross} we show the $\kappa$-momentum space and the mass-shells given by the above constraint.
In the massless case, using the relations above, it can be shown that the mass-shell condition reads
\be\label{func-C}
\mathcal{C}\left(1+\frac{\ell^2 \mathcal{C}}{4}\right)=0,
\ee
where $\mathcal{C}$ is the Casimir invariant of the $\kappa$-Poincar\'e algebra in bi-crossproduct coordinates:
\be
\mathcal C =- \frac{4}{\ell^{2}}\sinh^{2}(\ell E/2)+e^{\ell E} p^{2}.
\ee
Looking at the mass-shell, it is clear that one has to perform a further restriction on the allowed range for the embedding coordinates in order for the theory to be relativistic. In fact, a crucial request is that any mass-shell is completely within the allowed portion of de Sitter momentum space. Failure to meet this condition would result in the possibility for a  finite boosts to bring outside the allowed region some points that were originally within it. The restriction one has to enforce is given by the conditions
\begin{equation}
P_0>0, \quad P_4<0.
\end{equation}
However let us mention that at a field theoretic level the Hopf algebraic structures of the $\kappa$-Poincar\'e algebra ensure that the model is fully consistent without the further restriction above \cite{Arzano:2009ci}.

\subsection{Running of Hausdorff dimension of momentum space for the   $\kappa$-momentum space scenario}

In~\cite{dsrrsd} we showed that the $\kappa$-momentum space is characterized by a running of its Hausdorff dimension when going from the IR regime to the UV. We considered a general $D+1$ de Sitter manifold, and we allowed for the mass-shell to be a generic function of the $\kappa$-Poincar\'e Casimir, parameterized as $m^2=\mathcal{C}\left(1+\ell^{2\gamma} \mathcal{C}^{\gamma}\right)$.  Here we review the argument found in \cite{dsrrsd} for UV dimensional running, specializing to the $D=3$, $\gamma=1$ case, which is the one discussed in the previous subsection (the exact coefficient of the UV-dominant term in the mass-shell relation does not affect the UV value of the Haussdorf dimension). In doing so, we will add some remarks that will facilitate comparison with the AdS constructions.

The phenomenon of dimensional running can be characterized by choosing a set of ``linearizing coordinates", rendering the dispersion relations trivial in the UV, and examining the dimensionality associated with the integration measure in such coordinates.  The linearizing coordinates for $\cal C$, are~\cite{dsrrsd}:
\bea\label{lin0a}
\tilde E&=&\frac{2 \sinh (\ell E/2)} {\ell}\nonumber \\
\tilde p_i&=& p_i e^{ \ell E/2}.
\eea
which in the UV limit (defined as
$ E\rightarrow \infty$ and $ p\rightarrow 1/{\ell}$)
become:
\bea\label{lin0a1}
\tilde E&\approx &\frac{e^{\ell E/2}} {\ell}\nonumber \\
\tilde p_i&\approx& \frac{ e^{ \ell E/2}}{\ell}
\eea
(we note that in the UV limit $\tilde E\approx \tilde p$, even off-shell).
In terms of the new coordinates the measure (\ref{measurep}) is given by:
\be
d\mu= \tilde E ^{2}\tilde p^{2} d\tilde E\, d\tilde p\,.
\ee
As explained in~\cite{dsrrsd}, for on shell relations which in the UV limit take the form ${\cal C}^{1+\gamma}$, one finds
\be
d_H=\frac{6}{1+\gamma}\; ,
\ee
so in the case of interest here ($\gamma=1$) one finds that the Hausdorff dimension runs to 3 in the UV.

Notice that we could obtain the same result by linearizing directly the on-shell relation that comes out of the $\kappa$-momentum space construction, Eq.~(\ref{mass-shell}). This
amounts to choosing the embedding coordinates themselves as linearizing coordinates. In the UV, their relation to the bi-crossproduct coordinates is:
\bea\label{lin0b}
\tilde E&=&P_0\approx\frac{e^{\ell E}}{2\ell}(1+\ell ^2 p^2)  \nonumber \\
\tilde p_i&=&P_i \approx p_i e^{\ell E}
\eea
where the approximate signs refer to the UV limit approximation. We note also here that in the UV limit $\tilde E\approx \tilde p$, even off-shell.
The measure (\ref{measurep}) in the new coordinates now reads:
\be
d\mu=\frac{\tilde p^{2}}{\tilde E}d\tilde p\, d\tilde E
\ee
from which we can directly read $d_H=3$\footnote{In the more general $D+1$-dimensional case, and allowing for redefinitions of the mass-shell with UV limit $m^2=\ell^{2\gamma} (P_0^2-\vec P^2)^{1+\gamma}$ one would get \be\label{dhds1}
d_H=\frac{D}{1+\gamma}.
\ee
Note that this is another example of correspondence between the UV Hausdorff dimension of momentum space and the UV limit of the spectral dimension. The second one was computed in \cite{MicheleTomasz} for the 4- and 3- dimensional cases, and the results are in agreement with formula (\ref{dhds1}).}.
The last description will be useful in establishing a comparison with AdS constructions. It implies that if we take the MDR that comes most naturally out of dS (i.e., Eq.~(\ref{mass-shell})) then we would observe dimensional reduction from $D+1$ to $D$. This can be equivalently obtained from ${\cal C}$ with $\gamma=1$.

\section{AdS momentum space}
As with dS space, AdS momentum space can be described as a four-dimensional hyper-surface embedded in a five-dimensional flat space, this time with signature
$-,-,+,+,+$. The sub-manifold is now defined by:
\begin{equation}\label{adsdef}
  -P_0^2 + P_1^2 + P_2^2 + P_3^2 -P_4^2= -\frac{1}{\ell^2}\,
\end{equation}
and the corresponding line element is:
\be
ds^2=-dP_0^2 + dP_1^2 + dP_2^2 + dP_3^2 - dP_4^2 .
\ee
\subsection{Cosmological coordinates for AdS}\label{cosmo}
\begin{figure}
\includegraphics[scale=0.5]{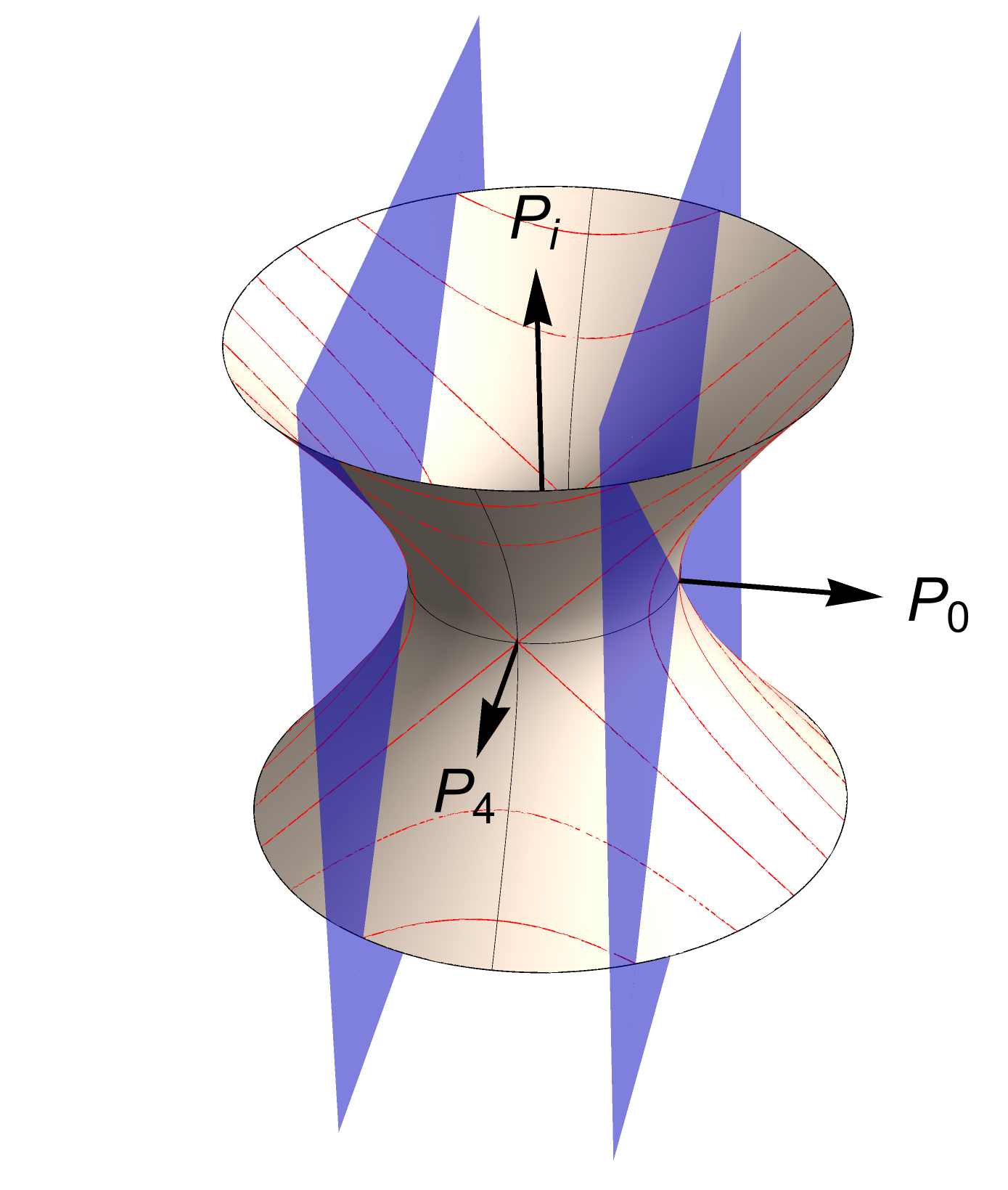}
\caption{Portion of AdS momentum space covered by cosmological coordinates. The condition for the allowed region is $-1/\ell<P_0<1/\ell$, which is the portion of the adS manifold between the two blue planes. The red lines represent the mass-shells, defined by the condition $P_4=const$.}
\label{fig:adSCosmological}
\end{figure}

In analogy with the $\kappa$-momentum-space construction over dS we seek coordinates casting a portion of AdS in the form of a cosmological metric (which is no longer a ``flat slicing'', as it was for dS). It can be shown (see Appendix A) that the cosmological AdS coordinates are defined by the following relation with the embedding coordinates:
\begin{eqnarray}\label{eq:AdSCosmo}
 {P_0}(E, \vec{p}) &=&  \frac{1}{\ell}\sin
\ell{E}\nonumber\\
 P_1(E, \vec{p}) &=&  p_1 \cos (\ell E) \nonumber\\
 P_2(E, \vec{p}) &=&  p_2 \cos (\ell E) \nonumber\\
 P_3(E, \vec{p}) &=&  p_3 \cos (\ell E)  \nonumber\\
 {P_4}(E, \vec{p}) &=& \frac{1}{\ell}\sqrt{1+(\ell p)^2}\cos\ell E
\end{eqnarray}
In these coordinates the metric reads:
\be\label{metradscosmo}
ds^2=-dE^2 + \cos^2 (\ell E){\left(\frac{dp^2}{1+\ell^2{p^2}} + p^2 d\Omega^2 \right)}.
\ee
The sub-manifold covered by these coordinates is defined by the constraint $-1/\ell\le P_0\le 1/\ell$, or, if we require the energy to be positive, $0\le P_0\le 1/\ell$. Using the line element (\ref{metradscosmo}) we easily deduce the invariant integration measure for AdS momentum space in these coordinates:
\be
d\mu(E,\ p)=\frac{\cos^3 (\ell E)}{\sqrt{1+\ell^2 p^2}} p^{2} dE dp.
\ee
In analogy with dS, the mass-shell relation can be inferred by imposing $P_4=const$ upon the surface condition:
\be
  -P_0^2 + \vec{P}^2=-\frac{1}{\ell^2}+P_4^2=-m^2\, \label{eq:adSCosmologicalMassShell}
\ee
From this we see that we must require that $m\le1/\ell$.
In terms of the cosmological AdS cordinates the mass-shell condition takes the form:
\be\label{casimirads}
-\frac{1}{\ell^2}\sin^2\ell E + p^2 \cos^2(\ell E)=-m^2.
\ee
In Figure \ref{fig:adSCosmological} we plot the sub-manifold of adS  covered by cosmological coordinates as well as the mass-shells given by the  constraint (\ref{eq:adSCosmologicalMassShell}).

\subsubsection{Maximal energy and speed of light in the UV limit}

\begin{figure}
\includegraphics[scale=0.8]{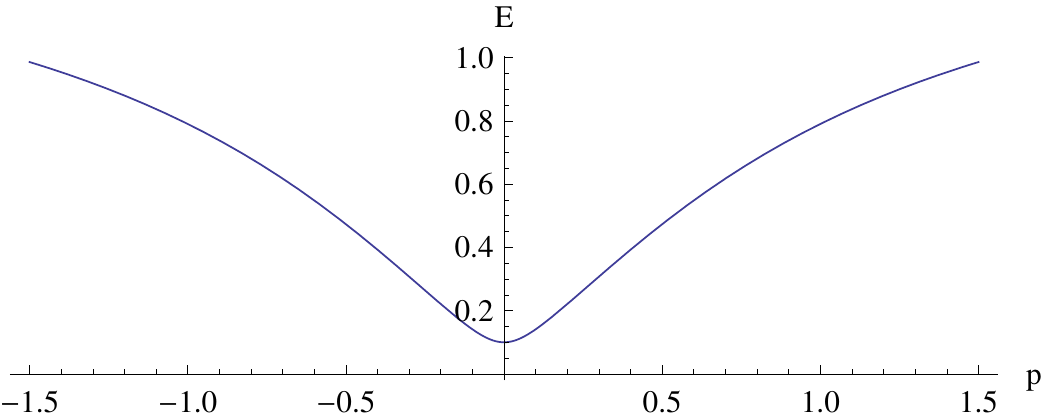}
\caption{Dispersion realation for a massive particle (\ref{eq:adSCosmoMDR}) in AdS momentum space with cosmological coordinates, with $\ell=1$ and $m=0.1$.}
\label{fig:adSCosmoMDR}
\end{figure}

The mass-shell condition given by Eq.~(\ref{casimirads}) implies the presence of a maximal energy in the theory (just like on the $\kappa$-momentum space there is a maximal spatial momentum). Let us consider a massless particle,
in this case  the dispersion relation is given by
\be
\frac{1}{\ell}\tan (\ell E)=p
\ee
and it is evident that
\be
E\le E_{max}=\frac{\pi}{2 \ell},
\ee
whereas there is no maximal spatial momentum. In addition, we see that the speed of light goes to zero as $p\rightarrow\infty$
\be
c=\frac{dE}{dp}\rightarrow 0.
\ee
This is nothing but the Carroll limit \cite{CarrollLimit}. For massive particles the mass shell can be written as:
\be
\cos^2\ell E=\frac{1-\ell^2 m^2}{1 +\ell^2 p^2}. \label{eq:adSCosmoMDR}
\ee
(remember that the constraint $m<1/\ell$ must be satisfied).
As $p\rightarrow\infty$ again we get $E\rightarrow E_{\max}$. Notice that if $m=1/\ell$ then the MDR does not fix the momentum, and the energy saturates.  A plot of the behaviour of the MDR is shown in Fig. \ref{fig:adSCosmoMDR}.

\subsubsection{Violation of Lorentz invariance}
Despite the care taken not to introduce a preferred frame, this has in fact sneaked in by virtue of the fact that the boundary of the sub-manifold is not invariant under the action of the Lorentz group. This is a crucial difference between the analogous 
constructions for dS (for which the boundary is $P_0=P_4$) and AdS (where the boundary is $|P_0|=1/\ell$). It might seem a subtle point, but the implications are obvious if we write down the Lorentz transformations in momentum space.

These can be described as 
a non-linear representation of the Lorentz group, inferred from the standard ones as applied to the embedding coordinates through the relations (\ref{eq:AdSCosmo}).

For a finite boost in the $\hat 1$ direction, the  explicit transformation rules are:
\bea
E'&=&\frac{1}{\ell}\arcsin\left( \gamma\, \ell\left(\frac{1}{\ell}\sin(\ell E)-vp_1\cos(\ell E)\right)\right)\\
p_1'&=&\frac{\gamma\left(p_1\cos(\ell E)-v\frac{1}{\ell}\sin(\ell E)\right)}{\sqrt{1-\ell^2\gamma^2\left(\frac{1}{\ell}\sin(\ell E)-vp_1\cos(\ell E)\right)^2}}\\
p_2'&=&\frac{p_2\cos(\ell E)}{\sqrt{1-\ell^2\gamma^2\left(\frac{1}{\ell}\sin(\ell E)-vp_1\cos(\ell E)\right)^2}}\\
p_3'&=&\frac{p_3\cos(\ell E)}{\sqrt{1-\ell^2\gamma^2\left(\frac{1}{\ell}\sin(\ell E)-vp_1\cos(\ell E)\right)^2}}
\eea
These can be shown to be generated by:
\be
L_{01}=p_1 \cos(\ell E) \partial _E +\frac{1}{\ell}\tan(\ell E) \partial_{p_1}.
\ee
It is obvious that there is something pathological with the finite transformations: as the transformation for the energy shows, there is clearly a maximal boost parameter, such that any larger boost would bring the value of energy outside the allowed range.

Another way to see that this framework is not invariant under finite boosts, is by noticing that any mass shell that goes through the allowed region of the adS manifold is not completely included within that region (see Fig. \ref{fig:adSCosmological}:), and there is no way to further restrict the allowed region so to solve the problem. This means that for any value of the mass and energy, there will always be a finite boost pushing the particle outside the allowed range of parameters.

Yet another indicator of the breakdown of relativistic invariance is the fact that not only is the energy $E$ bounded, but also the embedding one:
\be
P_0\le P_{0\,\max}=\frac{1}{\ell}
\ee
and this is clearly incompatible with its standard transformation rules under boosts. This is in sharp contrast with the corresponding situation in de Sitter space, where $p$ is bounded, but not its embedding counterpart.

\subsection{Horospherical coordinates}\label{horo}
An AdS coordinate system which mimics more closely the dS properties of the $\kappa$-momentum space is given by the so-called {\it horospherical coordinates} \cite{VileK,Lu:1996rhb}. In terms of the embedding coordinates they read
\bea
P_0 & = &\frac{1}{\ell}\cosh\left(\ell k_0\right) + \frac{\ell}{2} e^{\ell k_0} k_ik^i\,, \nonumber\\
P_4 & = &e^{\ell k_0} k_1\,, \nonumber\\
P_2 & = &e^{\ell k_0} k_2\,, \nonumber\\
P_3 & = &e^{\ell k_0} k_3\,, \nonumber\\
P_1 & = &\frac{1}{\ell} \sinh\left(\ell k_0\right) - \frac{\ell }{2} e^{\ell k_0} k_ik^i\,,
\eea
where now $k_i k^i = -k_1^2+k_2^2+k_3^2$. It is easy to verify that they satisfy
constraint (\ref{adsdef})
but only cover the $P_0 + P_1 > 0$ region, half AdS (see \cite{Lu:1996rhb}). The {\it spurious} embedding coordinate has to be time-like in this case and indeed it is easily verified that this must be $P_0$ since it diverges for $\ell\rightarrow 0$, the flat momentum space limit.

\begin{figure}
\includegraphics[scale=0.5]{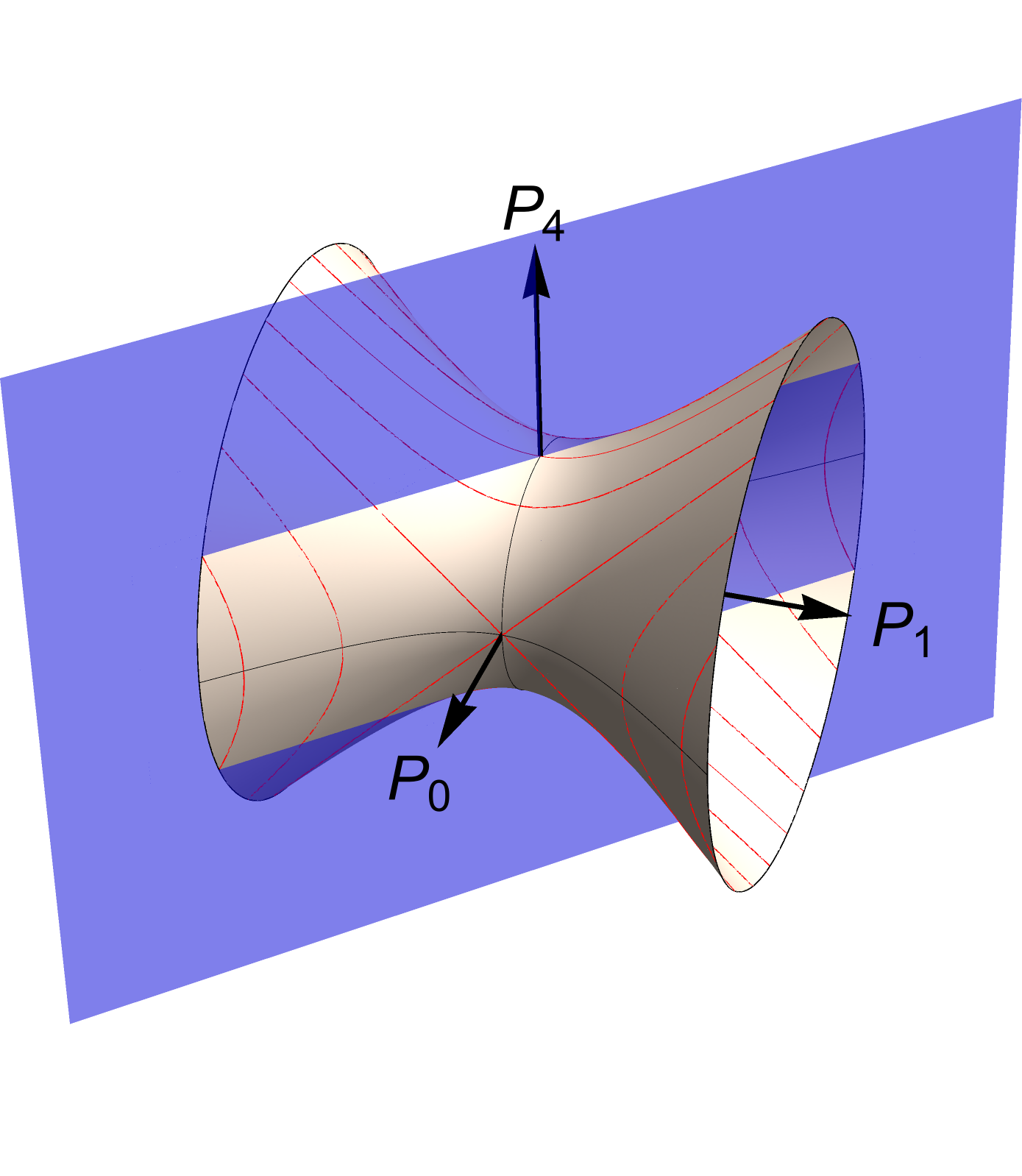}
\caption{Portion of AdS momentum space covered by horospherical coordinates. The condition for the allowed region is $P_0 + P_1 > 0$, which is the portion of the AdS manifold on the lower-right side of the blue plane. The mass-shell is given by $P_0=const.$ and is in red. Remember that now the $P_4$ coordinate is the one related to energy and $P_1$ is a spatial coordinate.}
\label{fig:adSHorospherical}
\end{figure}

Let us note that $k_0$ is now one of the components of the spatial momentum, and $k_1$ is the energy.  It is physically more transparent to write the new coordinates as:
\bea
P_0 & = &\frac{1}{\ell}\cosh\left(\ell p_1\right) + \frac{\ell}{2} e^{\ell p_1} (-E^2 + p_2^2 + p_3^2)\,, \nonumber\\
P_4 & = &e^{\ell p_1} E\,, \nonumber\\
P_2 & = &e^{\ell p_1} p_2\,, \nonumber\\
P_3 & = &e^{\ell p_1} p_3\,, \nonumber\\
P_1 & = &\frac{1}{\ell} \sinh\left(\ell p_1\right) - \frac{\ell }{2} e^{\ell p_1} (-E^2 + p_2^2 + p_3^2)\,.
\eea
In such coordinates the line element is given by
\be
ds^2=e^{2\ell p_1} (-dE^2 + dp_2^2 + dp_3^2) + dp_1^2\,,
\ee
and the associated integration measure is:
\be
d\mu =e^{3\ell p_1}\, dE \,dp\,.
\ee
Analogously to what we have done in the previous subsections, we find the mass-shell condition by requiring that the spurious time-like coordinate is constant, which in this case amounts to asking $P_0=const$:
\be
P_0^2-\frac{1}{\ell^2}=-\vec P^2+P_4^2=-m^2.
\ee
Also here this implies that the mass can not be arbitrarily large, $m\le \frac{1}{\ell}$. In terms of the embedding coordinates the mass-shell condition reads:
\bea
-m^2&=&-e^{2\ell p_1} E^2+e^{2\ell p_1} (p_2^2 +p_3^2)\nonumber\\
&&+{\left(\frac{1}{\ell}\sinh (\ell p_1)-\frac{\ell}{2}
e^{\ell p_1} (-E^2 + p_2^2 +p_3^2)\right)}^2\,\,\,\, \label{eq:CasimirHorospherical}
\eea
In Figure \ref{fig:adSHorospherical} we plot the sub-manifold of AdS  covered by horospherical coordinates, as well as the mass-shells.
Despite the obvious anisotropy introduced by these coordinates, a deformed description of rotations exists, ad can be derived in analogy of what was done for boosts in cosmological coordinates. This means that the sub-manifold is invariant under infinitesimal transformations. However, one can see that the further restriction of the manifold to the $P_1>0$, $P_0>0$ region has to be enforced in order to have worldlines not exiting the manifold (i.e. to have invariance under finite boosts). This in turn leads to a breakdown of invariance under rotations, because of the condition $P_1>0$ due to the fact that embedding coordinates transform according to the standard Lorentz transformations.

\subsubsection{Anisotropic speed of light}
It is interesting to look at the behaviour of the speed of light in this model.  For notational simplicity we restrict to the case of $2+1$-dimensional AdS momentum space and we write the spatial momenta in polar coordinates $p_1=p \,\cos \theta, p_2=p\, \sin \theta$.  The dispersion relation for a massless particle reads
\be
\begin{split}
&E^2=\frac{1}{2\ell^2}\Bigg(2 e^{-2\ell p \cos \theta}+  (2+\ell^2 p^2 \sin^2 \theta)-e^{-\ell p \cos \theta}\cdot\\
&\cdot\sqrt{   16-p^2\ell^2 \sin^2 \theta \Big( 4+e^{2\ell p \cos \theta} (4+3\ell^2 p^2 \sin^2 \theta) \Big) }\Bigg)\, .
\end{split}
\ee
The deformation of rotations which guarantee local invariance of the manifold leads to a direction-dependent dispersion relation and as a consequence to a direction-dependent speed of light.

\begin{figure}
\includegraphics[scale=0.7]{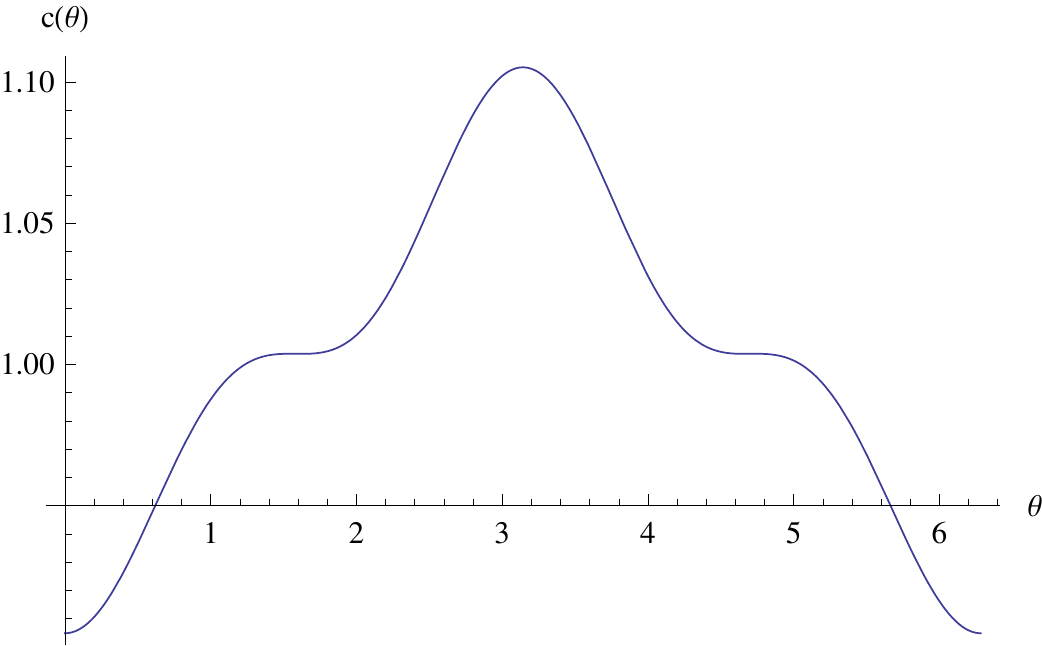}
\caption{Speed of light in horospherical coordinates as a function of $\theta$, for $\ell=1, p=0.1$.}
\label{fig:horosphericalspeedoflight}
\end{figure}

The general expression is quite complicated (a plot of its angular dependence can be seen in Fig. \ref{fig:horosphericalspeedoflight}), so here we only write down the two special cases $\theta=0$ (speed of light along the $p_1$ direction) and $\theta=\pi/2$ (speed of light along the $p_2$ direction)
\bea
c(p,\theta=0)&=&e^{-\ell p} \\
c(p,\theta=\pi/2)&=&\frac{\ell p \Big( 4+3 \ell^2 p^2+A \Big)}{A\sqrt{8+2\ell^2 p^2-2A}}
\eea
where $A=\sqrt{16-8\ell^2 p^2-3\ell^4 p^4}$.
Note that the speed of light becomes imaginary whenever $p$ and $\theta$ are such that the condition $P_1>0$ is violated. This makes it impossible to reach infinite speed of light in any direction.  Strange as this model might be it could be a good framework for encoding fundamental anisotropy, with constraints on it encapsulating observational facts, like those derived from modern day versions of the Michelson-Morley experiment. Occasional claims for cosmological anisotropy (e.g.~\cite{evil})  could also be embedded in this model, but the
matter is beyond the scope of this paper.

\subsection{``Static'' coordinates}\label{static}
Static coordinates cover the full AdS manifold, and so clearly they do not break Lorentz invariance, but rather deform it. They can be defined from the embedding coordinates via: 
\begin{eqnarray}\label{static-ads}
P_0 &=&\frac{1}{\ell}\sin(\ell E) \cosh (\ell p') \nonumber\\
P_r &=&\frac{1}{\ell}\sinh (\ell p') \nonumber\\
P_4 &=&\frac{1}{\ell}\cos(\ell E) \cosh (\ell p')
\end{eqnarray}
where $P_r\equiv\sqrt{P_1^2+P_2^2+P_3^2}$.
The line element in these coordinates is:
\be
ds^2=-\cosh^2(\ell p')dE^2+ dp'^2 + \frac{1}{\ell^2}\sinh^2(\ell p')d\Omega.
\ee
We can also use an areal coordinate:
\be
p= \frac{\sinh(\ell p')}{\ell}
\ee
resulting in:
\begin{eqnarray}\label{static-ads1}
P_0 &=&\frac{1}{\ell}\sin(\ell E) \sqrt{1+(\ell { p})^2}\nonumber\\
P_r &=& { p} \nonumber\\
P_4 &=&\frac{1}{\ell}\cos(\ell E) \sqrt{1+(\ell {p})^2}
\end{eqnarray}
for which the line element is:
\be
ds^2=-(1+(\ell { p})^2)dE^2+ \frac{d{ p}^2}{1+(\ell { p})^2} + { p}^2d\Omega.
\ee
Notice that invariance under (deformed) Lorentz transformations is not spoiled if we ask the energy to be positive, i.e. this requirement is compatible with the deformed transformation rules. This can be seen from (\ref{static-ads1}): enforcing the positivity of $E$ is equivalent\footnote{In this work we always assume $\ell>0$.} to enforcing the positivity of $P_0$. But the embedding coordinate $P_0$ transforms with the standard Lorentz transformations, so asking it to be positive works in the same way as in the usual special relativistic case.  Another interesting feature of such coordinates is that the integration measure is undeformed:
\be
d\mu={ p}^{2} d{ p}\, dE.
\ee
We can find the mass-shell relation again by requiring that $P_4=const$:
 \be
  -P_0^2 + \vec{P}^2=-\frac{1}{\ell^2}+P_4^2=-m^2\,,
\ee
which in static coordinates becomes:
\be
-\frac{\sin^2(\ell E)}{\ell^2}[1+(\ell{ p})^2] +{p}^2=-m^2.
\ee
In order to explore the physics and UV limit we consider massless particles. We see that their spatial momentum is unbounded, but their energy tends to a
maximum:
\bea
{ p}&\rightarrow&\infty\\
E&\rightarrow& E_{max}=\frac{\pi}{2\ell}.
\eea
We take these limiting values as the UV limit of the model. The speed of light is given by:
\be
c=\frac{dE}{d{p}}=\frac{1}{1+(\ell { p})^2}
\ee
and in the UV limit this goes to zero, what is known in the literature as the {\it Carroll limit} \cite{CarrollLimit}.

\section{UV Dimensional reduction in AdS momentum space}\label{run}
In a series of recent papers~\cite{measure,dsrrsd} we have shown that it is possible to characterize the phenomenon of dimensional reduction in the UV dispensing with the concept of spectral dimension altogether. This is beneficial, as the latter appeals to a fictitious time parameter, requires the Euclideanization of the space, and is not always based on a properly defined probability distribution. Instead we showed that we could transfer all the non-trivial effects of the MDRs into the measure, adopting linearizing variables and then study the Hausdorff dimension of the energy-momentum space in these variables. This is a physically clearer procedure, and asymptotically (i.e. in the deep UV limit) it produces results coinciding with those using the spectral dimension in all known cases. Given the difficulties in defining asymptotic spectral dimension for AdS momentum space, we favour our procedure here.

As with~\cite{measure,dsrrsd} we shall be concerned with MDRs which in the UV limit have the form:
\be
\Omega= f(\mathcal C) \approx \mathcal C^{1+\gamma}
\ee
where $\mathcal C$ is the Casimir invariant of the theory. We will examine the UV running of the Hausdorff dimension for cosmological and static coordinates which have a clear UV limit leaving aside the case of horospherical coordinates whose anisotropic nature renders the notion of UV limit ambigous. 

\subsection{Cosmological coordinates}
Following~\cite{measure,dsrrsd}, we find linearizing variables in 2 steps: first by assuming $\gamma=0$, then generalizing to $\gamma\neq 0$.  When $\gamma=0$ (i.e. when the MDRs are just the Casimir) the linearizing coordinates are just the embedding coordinates, as in (\ref{eq:AdSCosmo}): $\tilde E\equiv P_0, \tilde p\equiv P$. In $D+1$ space-time dimensions the momentum space integration measure in such variables is:
\be\label{linmeasure}
d\tilde\mu=\frac{\tilde p^{D-1}}{\sqrt{1+\ell^2(\tilde p^2-\tilde E^2)}}d\tilde Ed\tilde p,
\ee
which, in the UV limit (as defined above), becomes:
\be
d\tilde\mu\approx \tilde p^{D-2}d\tilde Ed\tilde p.
\ee
Therefore the Hausdorff dimension is reduced by 1. We note that this is just the general measure studied in \cite{dsrrsd}:
\be
d\mu(\tilde E,\tilde p)\propto \tilde p^{D_x-1}  \tilde E^{D_t-1} d\tilde Ed\tilde p
\ee
with values $D_t=1$, and $D_x=D-1$.

If $\gamma\neq 0$ the linearizing coordinates can be found by following the ``step 2''  described in~\cite{dsrrsd}, but we stress that the procedure here does not rely on Euclideanization. In~\cite{dsrrsd} we were dealing with an Euclideanized version of momentum space (because we wanted to study the {\it asymptotic} coincidence of spectral and Hausdorff dimensions), but the procedure carries through with a minimal adaptation if we remain Lorentzian. All we need do is introduce hyperbolic (instead of spherical) polar coordinates:
\bea
\tilde E&=&r\cosh \theta\\
\tilde p&=&r\sinh \theta
\eea
so that the MDRs become
\be
\Omega = r^{2(1+\gamma)}.
\ee
We can then define a linearizing variable
\be
\hat r= r^{1+\gamma}
\ee
such that:
\be
d\mu \propto \hat r^{\frac{D_t+D_x}{1+\gamma}-1}(\cos\theta)^{D_t-1}(\sin\theta)^{D_x-1}\, d\hat r\, d\theta
\ee
leading to the conclusion that in the UV:
\be
d_H=\frac{D_t+D_x}{1+\gamma}.
\ee
For the AdS model we are considering this therefore becomes:
\be\label{dhadscosmo}
d_H=\frac{D}{1+\gamma}.
\ee

\subsection{Static coordinates}
Similarly to what happens for the cosmological coordinates, the linearising coordinates for the model are the embedding coordinates found in (\ref{static-ads1}).
The integration measure is the same as (\ref{linmeasure}). However the UV limit now entails $\tilde E\approx \tilde p$ leading to an undeformed measure. This model therefore has non-trivial physical effects (e.g. it has a Carroll limit) but it does {\it not} present running of the dimensionality, if $\gamma=0$. If $\gamma\neq 0$ one can straightforwardly calculate 
\be\label{dhadsstat}
d_H=\frac{1+D}{1+\gamma}\, ,
\ee
and we therefore have a non-trivial running of the dimensionality.

\section{de Sitter momentum space in static coordinates}\label{desitter}
The first two models above arise from attempts to construct four-momenta defined on AdS space using a {\it duality} approach to dS space of momenta associated to $\kappa$-Poincar\'e.  The third model based on ``static coordinates", however, was proposed without reference to a dS construction, so one might wonder what the equivalent dS model would be. As we shall see, whilst static coordinates lead to an AdS momentum space model which does not break any symmetry, its dS counterpart breaks Lorentz invariance.

Static coordinates for dS may be built from:
\begin{eqnarray}\label{static-ds}
P_0 &=&\frac{1}{\ell}\sinh(\ell E) \sqrt{1-(\ell { p})^2}\nonumber\\
P_r &=& { p} \nonumber\\
P_4 &=&\frac{1}{\ell}\cosh(\ell E) \sqrt{1-(\ell { p})^2}
\end{eqnarray}
leading to metric:
\be
ds^2=-(1-(\ell { p})^2)dE^2+ \frac{d{ p}^2}{1-(\ell { p})^2} + { p}^2d\Omega.
\ee
and an undeformed integration measure. The Casimir is:
\be
{\cal C}=-\frac{\sinh^2(\ell E)}{\ell^2}[1-(\ell{ p})^2] +{ p}^2=m^2.
\ee
and we see that the theory has a maximum spatial momentum, $ p_{max}=1/\ell$, but unbounded energy, just like the $\kappa$-Poincar\'e case. (This maximal momentum coincides with the location of de Sitter's horizon, in the counterpart position space version of the space.)The UV limit may be accordingly defined by:
\bea
{ p}&\rightarrow&{ p}_{max}=\frac{1}{\ell}\\
E&\rightarrow& \infty.
\eea
and in such limit the speed of light
\be
c=\frac{dE}{d{ p}}=\frac{1}{1-(\ell p)^2}
\ee
goes to infinity in the UV.

All of these features are very similar to what is found in $\kappa$-Poincar\'e in the {\it bicrossproduct basis}. However this model breaks Lorentz symmetry in a way
that mimics closely what happens for cosmological AdS coordinates. Indeed looking at the embedding coordinates  (\ref{static-ds}), it is clear that the maximum value of momentum is reflected into a maximum value of the embedding coordinate $P_r$.
Since the embedding coordinates transform according to standard Lorentz transformations, this is inconsistent with the relativity of inertial frames.

We conclude by noticing that this model also exhibits running of dimensionality. Working out the integration measure in linearizing coordinates, $\tilde E\equiv P_0, \;\tilde p\equiv P_r$, leads to:
\be\label{linmeasuredS}
d\tilde\mu=\frac{\tilde p^{D-1}}{\sqrt{1-\ell^2(\tilde p^2-\tilde E^2)}}d\tilde Ed\tilde p\,,
\ee
that in the UV limit becomes
\be
d\tilde\mu\approx \frac{\tilde p^{D-1}}{\tilde E}d\tilde Ed\tilde p\,.
\ee
Following our standard calculation, we find for the UV Hausdorff dimension of momentum space
\be\label{dhdsstat}
d_H=\frac{D}{1+\gamma}\; .
\ee
This is suggestively similar to what we found for AdS in cosmological coordinates. It also matches the result obtained for dS linearising directly from the embedding coordinates, as discussed in Section~\ref{dSbackground}.

\section{Conclusions}\label{concs}
In this paper we took a first stab at defining a curved momentum space based on AdS geometry, in analogy with previous work for dS space. A number of significant novelties were uncovered in the process.

The equivalent of the bicrossproduct basis was sought in two ways. Firstly, we noted that we can regard the bicrossproduct basis as the momentum space counterpart of the ``cosmological'' covering of dS, and sought similar coordinates for AdS. We found that the equivalent construction for momentum space AdS, while simpler than dS and superficially more elegant, in fact breaks Lorentz invariance instead of deforming it. The model must introduce a preferred frame because the boundary of the corresponding sub-manifold is no longer invariant under the action of the Lorentz group. The ensuing model may thus be useful as a way of encoding subtle frame-dependence due to the boundary effects: the frame dependence is only obvious with sufficiently large Lorentz transformations.

One can also look at the bicrossproduct momenta associated to $\kappa$-Poincar\'e as {\it horospherical coordinates} on dS momentum space. Such coordinates can be introduced also for AdS momentum space and we defined the associated energy and momentum. We find that the corresponding construction introduces spatial anisotropy in momentum space and thus one must deform not only Lorentz symmetry but also spatial rotations. However, similarly to what happens in the ``cosmological" coordinates setting, the boundary is not invariant under rotations, breaking isotropy. The result is awkward in several other ways, including the fact that the speed of light and the MDRs are anisotropic. Again this may serve as a useful way of encoding phenomenology, in this case anisotropic dispersion relations and anisotropic speed of light.

A third construction, based on ``static coordinates'', whilst not mimicking the usual set up for $\kappa$-Poincar\'e space, proves to be the best one conceptually and in terms of simplicity. It leads to an undeformed integration measure and a very simple Casimir invariant. It models a maximal energy and unbounded spatial momentum without introducing a preferred frame. The speed of light goes to zero in the UV limit, and this is achieved isotropically. We advocate this construction as the most conservative model for AdS momentum space.  For completeness, in this paper we have also considered a dS model based on static coordinates, the counterpart to the last AdS model proposed in this paper. Curiously the dS static model breaks Lorentz invariance in a way similar to what happens to the AdS model in cosmological coordinates.

As a first application of these models we investigated the phenomenon of running of the dimensionality.  We did this by considering ``linearizing'' coordinates (i.e. coordinates which render the dispersion relations trivial) and evaluating the integration measure in terms of them, to find the associated Hausdorff dimension. This procedure was considered in the past~\cite{measure,dsrrsd}, and found to match the spectral dimension in the UV limit in all cases studied.  In this paper we found that the (Lorentz breaking) AdS model based on cosmological coordinates runs to:
 \be
d_H=\frac{D}{1+\gamma}
\ee
in the UV limit, whereas the (non-Lorentz breaking) model based on static coordinates runs to:
\be
d_H=\frac{1+D}{1+\gamma}
\ee
showing further that the two models are physically distinct models.  It is curious that the equivalent result for dS in static coordinates matches the result found for AdS in cosmological coordinates (cf. Eq.(\ref{dhdsstat}) and Eq.(\ref{dhadscosmo})).  This also matches the result for the $\kappa$-Poincar\'e space \cite{MicheleTomasz} if we linearize the Casimir coming directly from the embedding variables, as explained in the discussion leading to Eq.~(\ref{dhds1}). Could this be pointing us to an interesting duality?

\section{Acknowledgments}
We were all supported by the John Templeton Foundation. The work of MA was also supported by a Marie Curie Career Integration Grant within the 7th European Community Framework Programme. JM was also funded by a STFC consolidated grant and the Leverhulme Trust.

\appendix
\section{Derivation of cosmological coordinates for AdS}
One starts by setting~\cite{bengtsson}:
\bea
P_0&=&\frac{1}{\ell}\sin\ell{E}\nonumber\\
P_i&=&{\hat P}_i \cos\ell{E}\nonumber\\
P_4&=&{\hat P}_4 \cos\ell{E}
\eea
where $i=1,2,3$ and $P_\mu$ are the embedding coordinates. In this way Eq.(\ref{adsdef}) becomes a condition requiring the spatial homogeneous leaves to be hyperboloids:
\be\label{constraint}
{\hat P}_i^2-{\hat P}_4^2=-\frac{1}{\ell^2} .
\ee

In terms of these coordinates the metric induced on the 4-surface is the cosmological rendition of (a portion of) AdS:
\be
ds^2=-dE^2+\cos^2(\ell E)d\sigma^2
\ee
where the spatial metric is
\be
d\sigma^2=d{\hat P}_1^2+d{\hat P}_2^2+d{\hat P}_3^2-
d{\hat P}_4^2
\ee
subject to (\ref{constraint}).
Introducing polar coordinates in the $\{ {\hat P}_i\}$ space:
\bea
{\hat P}_1&=&p\cos\theta\nonumber\\
{\hat P}_2&=&p\sin\theta\cos\phi\nonumber\\
{\hat P}_3&=&p\sin\theta\sin\phi
\eea
ensures that $p$ will be a comoving areal coordinate. Indeed, then
$d\sigma^2=dp^2+p^2d\Omega^2 -dP_4^2$, and $dP_4^2$ can at most correct the
$dp^2$ component of the metric. Specifically we can solve
(\ref{constraint}) as:
\be
{\hat P}_4=\frac{\sqrt{1+\ell^2p^2}}{\ell}
\ee
and by differentiating and inserting in $d\sigma^2$ we get the cosmological form of the AdS metric:
\be
ds^2=-dE^2 + \cos^2 (\ell E){\left(\frac{dp^2}{1+\ell^2{p^2}} + p^2d\Omega^2 \right)}.
\ee
The explicit expression relating the two sets of coordinates is therefore:
\begin{eqnarray}
 {P_0}(E, \vec{p}) &=&  \frac{1}{\ell}\sin
\ell{E}\nonumber\\
 P_1(E, \vec{p}) &=&  p \cos (\ell E) \cos\theta \nonumber\\
 P_2(E, \vec{p}) &=&  p \cos (\ell E) \sin\theta \cos\phi\nonumber\\
 P_3(E, \vec{p}) &=&  p \cos (\ell E) \sin\theta\sin\phi \nonumber\\
 {P_4}(E, \vec{p}) &=& \frac{1}{\ell}\sqrt{1+(\ell p)^2}\cos\ell E
\end{eqnarray}
This transformation can be abbreviated using notation:
\begin{eqnarray}\label{bicross-ads}
 {P_0}(E, \vec{p}) &=&  \frac{1}{\ell}\sin
\ell{E}\nonumber\\
 P_r(E, \vec{p}) &=&  p \cos \ell E \nonumber\\
 {P_4}(E, \vec{p}) &=& \frac{1}{\ell}\sqrt{1+(\ell p)^2}\cos\ell E
\end{eqnarray}
where the $\{ P_i\}$ are to be obtained from $P_r$ via the usual polar coordinate formulae. Then Eq.(\ref{bicross-ads}) is valid in any number $D$ of spatial dimensions, as long as we employ the standard polar coordinate $D-1$ angles. In these coordinates the AdS metric is:
\be\label{metrads}
ds^2=-dE^2 + \cos^2 (\ell E){\left(\frac{dp^2}{1+\ell^2{p^2}} + p^2d\Omega_{D-1}^2 \right)}.
\ee

\end{document}